\documentclass[preprint,prd,noshowpacs]{revtex4}
\usepackage{amssymb}
\usepackage{amsmath}
\usepackage{amsfonts}
\usepackage{graphicx}
\begin{document}

\title{Exact radial solution in 2+1 gravity with a real scalar field}

\author{Hans-J\"urgen  Schmidt}
\email{hjschmi@rz.uni-potsdam.de}
\affiliation{Institut f{\"u}r Mathematik, Universit{\"a}t Potsdam 
Am Neuen Palais 10, D-14469 Potsdam, Germany}
\author{Douglas Singleton}
\email{dougs@csufresno.edu}
\affiliation{Department of Physics, Institut Teknologi Bandung, Indonesia} 
\affiliation{Institut f{\"u}r Mathematik, Universit{\"a}t Potsdam 
Am Neuen Palais 10, D-14469 Potsdam, Germany}
\affiliation{Physics Department, CSU Fresno, Fresno, CA 93740-8031, USA}

\date{\today}

\begin{abstract}
In this paper we give some general considerations about circularly symmetric, static
space-times in $2+1$ dimensions,
focusing first on the surprising (at the time) existence of the BTZ black hole solution.
We show that BTZ black holes and Schwarzschild black holes in $3+1$ dimensions originate from
different definitions of a black hole. There are two by-products of this general
discussion: (i) we give a new and simple derivation of $2+1$ dimensional Anti-de Sitter (AdS) 
space-time; (ii) we present an exact solution to $2+1$ dimensional gravity coupled to a self-interacting 
real scalar field. The spatial part of the metric of this solution is flat but the temporal part 
behaves asymptotically like AdS space-time. The scalar field has logarithmic behavior
as one would expect for a massless scalar field in flat space-time. The solution can be compared
to gravitating scalar field solutions in $3+1$ dimensions but with certain 
oddities connected with the $2+1$ dimensional character of the space-time. The solution is unique 
to $2+1$ dimensions; it does not  carry over to $3+1$ dimensions. 
\end{abstract}

\maketitle


\section{Introduction}
\setcounter{equation}{0} 
The study of general relativity in dimensions larger than the $3+1$ that are observed 
has a history almost as long as general relativity itself. Already in \cite{kaluza} 
a $4+1$ dimensional space-time was investigated in an attempt to unify 
gravitational and electromagnetic interactions. In general, Kaluza-Klein 
theories \cite{wesson} and string theory \cite{gasperini} work with space-time 
dimensions greater than $3+1$. The study of general relativity in lower 
dimensions -- specifically $2+1$ dimensions -- has a more recent history. 
Starting in the mid-1980s several researchers launched investigations into $2+1$ 
gravity \cite{deser}, \cite{giddings}, \cite{jackiw}, \cite{hooft}, and \cite{witten}. A 
recent review of $2+1$  gravity can be found in \cite{carlip}. The motivation was to 
understand gravity in the  simpler context of $2+1$ dimensions in the hope that this 
would help shed some light on the more realistic but more  complex $3+1$ 
dimensional case -- in particular in regard to the question of quantizing gravity.  

There are some well known differences \cite{deser}, \cite{giddings} between
gravity in $2+1$ and $3+1$ dimensions. First, in the absence of sources, $2+1$ 
space-times are locally flat.  Related to this is the fact that the Weyl tensor vanishes
in $2+1$ dimensional gravity. Second, in $2+1$  dimensional gravity  there are 
no gravitational waves. Third, in $2+1$ dimensions there exists
no adequate Newtonian limit for the Einstein field equation  \cite{giddings}.
Finally, many geometric and topological properties of  spaces behave completely 
different in even dimensions in comparison  to an  odd number of  dimensions.
For example  \cite{hjs}, antipodal identification within the standard $n$-sphere 
$S^n$ leads  to an orientable factor space for odd values of $n$ only.  

One might surmise that since $2+1$ space-time is flat outside the presence of sources
that a point source in $2+1$ dimensions would have no effect except exactly at the 
location of the source. However, in \cite{deser} and  \cite{giddings}, single and 
multiple point sources were studied and it was found that a point source introduced
a conical singularity in the space-time. Thus although $2+1$ space-time was locally 
flat it would be globally conical with the deficit angle depending on the mass of the 
point source. Light rays in the presence of a  $2+1$ dimensional point mass, 
although moving locally in flat space-time, could be ``lensed" due to the global 
conical structure of the space-time. 

Given the requirement that the local presence of a mass-energy source was required 
to curve space-time it was a surprise that it was possible to construct a black 
hole space-time \cite{btz} (a review of $2+1$ dimensional BTZ black holes is given
in \cite{carlip-bh}). This was accomplished by considering a point mass
source in a space-time with a negative cosmological constant, $\Lambda <0$ -- i.e. 
Anti-de Sitter (AdS) space-time. Interpreting this negative cosmological constant
as an energy-momentum tensor $T_{\mu \nu} ^{AdS} \propto \Lambda g_{\mu \nu}$ 
would imply a negative energy density $\rho ^{AdS} <0$. We point this out to 
emphasize that sources in $2+1$ dimensions can be odd or anti-intuitive from 
a $3+1$ perspective. The exact solution we give below fits into this description 
since we will find that the curvature of the $2+1$-dimensional  space-time comes entirely
from a non-zero, radial pressure. 

As we will show in the following section there is nothing incompatible between the
statement in \cite{giddings} (see p. 759) that 
``Black holes do not exist in a 3-dimensional space-time." and 
the existence of the BTZ black hole solution. The resolution lies in the 
more restrictive definition of black holes used in \cite{giddings} as compared to
the usual manner in which the BTZ solution is characterized as a black hole solution. For the usual 3+1 
Schwarzschild black hole, as in many other cases, the horizon is defined by global 
considerations. However the horizon in these cases is located at exactly the same 
hypersurface as a Killing horizon, i.e. a hypersurface where one of the Killing vectors 
changes its character from space-like to time-like. This latter condition is defined locally, 
and is much easier to verify. For many space-times, like the 3+1 
dimensional Schwarzschild metric, the definition of the black hole via the Killing 
horizon is equivalent to the horizon defined via global considerations since these two 
definitions of horizon coincide in these cases. However, in the case of the 2+1 
dimensional BTZ space-time the situation is different since this metric does not 
possess a Killing horizon. In short: The BTZ black hole is locally static at all of
its points, whereas the Schwarzschild black hole is locally static outside the 
horizon only. We will give more mathematical details of this distinction in the following section.

After our general discussions of $2+1$ dimensional space-times we obtain a specialized but exact 
closed form solution of a gravitating, self-interacting, real scalar 
field in $2+1$ dimensions. To our knowledge not much work 
has been done in investigating gravitating  scalar fields in $2+1$ dimensions. There 
have been numerous studies of gravitating scalar fields  in $3+1$ dimensions. For
a complex scalar field in $3+1$ dimensions \cite{scalar31} both without as well 
as with self-interactions \cite{scalar31a} it was shown that one could find boson ``star" 
solutions -- localized, spherically symmetric configurations of the scalar field. Real 
scalar fields in $3+1$ dimensions were investigated in \cite{kodama},  
\cite{baekler}, and \cite{schmo}. In the case of a real scalar field it was not
possible to find a non-singular solution {\it except} in the case when one allowed the 
scalar field to be a phantom field \cite{kodama}, i.e. the kinetic energy term for the 
scalar field had a negative sign. However in this case of a phantom, real
scalar field, the topology of the solution became that of a wormhole. 
On another level,  non-singular solutions in $3+1$ dimensions  with a massive  
real scalar field as source  have been shown to exist in Einstein's theory 
within the set of closed Friedmann universe models, see e.g.  \cite{jetzer}.

There have been some studies of gravitating scalar fields in $2+1$ dimensions but
almost universally connected with time dependent situations such as collapse to a 
BTZ black hole \cite{hirschmann}. As far as we have found there has only been
one study of $2+1$ static solutions \cite{beato} similar to those investigated 
in $(3+1)$, see \cite{kodama}, \cite{baekler}, and  \cite{schmo}. This article 
presents one such static solution of a gravitating scalar field  in $2+1$.  
The exact solution presented here can be thought of as a localized particle-like 
solution in $2+1$ dimensions since the curvature scalars -- the Ricci 
scalar $R^\alpha _\alpha$ and Ricci tensor squared,
$R^{\alpha \beta} R_{\alpha \beta}$ -- behave like $1/r^2$ and $1/r^4$ 
respectively. The singularity in these scalars at $r=0$ can be thought of as the location 
of the scalar ``particle". Additionally the scalar field behaves like $\ln (r)$ which is
typical of  a massless field in $2+1$ dimensions in the absence of gravity with the 
source located at $r=0$. While the spatial part of our metric is flat (i.e. 
$dr^2 + r^2 d \theta^2$) the temporal part of  our metric 
behaves like the asymptotic form of AdS space-time (i.e. the temporal part of the metric 
is $K r^2 dt^2$ which is the same as the temporal part of AdS space-time in the 
limit $r \rightarrow \infty$). This can be contrasted with the BTZ 
solution which is asymptotically AdS space-time in both the time and radial coordinate.  

\section{General discussion of static, spherically symmetric $2+1$ dimensional space-times} 

In this section we will give a general discussion of static, rotationally symmetric 
metrics in $2+1$ dimensions. In the following section we will show that a specialized 
form of the general metric discussed in this section has an exact solution
when coupled to a real scalar field with a Liouville self-interacting potential.  The
form of the $2+1$ dimensional metric we begin with has the form
\begin{equation}
\label{2.1}
ds^2 = a^2(r) dt^2 - b^2(r)dr^2 - r^2 d \theta^2
\end{equation}
with $\theta=0$ and $\theta = 2 \pi$ being identified, if the circumferences of the 
circles $r = {\rm const.}$ are strictly monotonous functions of the radial coordinate. 
In this section, we take $x^0=t$, $x^1=r$, and $x^2=\theta$, and $i, j = 0,1, 2$.
We assume the functions $a(r)$ and $b(r)$ are positive, that $r \ge 0$, and that 
\begin{equation} 
\label{2.2}
\lim_{r \to 0} \ b(r)=1 
\end{equation}
to ensure that no conical singularity appears at $r=0$, the center of symmetry. 

The Ricci tensor, $R_{uv}$, for metric \eqref{2.1} in the coordinates chosen here, 
is diagonal. The three eigenvalues  $\lambda_0$, $\lambda_1$,  $\lambda_2$ of $R_{uv}$
(defined as $R_{ii} = \lambda_i  g_{ii}$ , where $\lambda_i$ is the eigenvalue into the
$x^i$-direction) are given by 
\begin{equation}  
\label{2.3}
\lambda_0= \frac{a''}{ab^2} - \frac{a'b'}{ab^3} + \frac{a'}{rab^2},
\end{equation}
\begin{equation} 
\label{2.4}
\lambda_1= + \frac{a''}{ab^2} - \frac{a'b'}{ab^3} - \frac{b'}{rb^3},
\end{equation}
\begin{equation}  
\label{2.5}
\lambda_2= - \frac{b'}{rb^3} + \frac{a'}{rab^2},
\end{equation}
where the dashes denote  $d/dr$. The curvature scalar, $R$, is given  by 
\begin{equation}  
\label{2.6}
R =  \lambda_0 + \lambda_1 +\lambda_2 .
\end{equation}
Let us look at the consequence if we require $\lambda_0 = \lambda_1$. From  eqs.  \eqref{2.3} and \eqref{2.4} 
this is equivalent to requiring  $a'/a = - b'/b$, i.e., that  the 
product $a \cdot b$ is a positive constant. A coordinate transformation 
in metric \eqref{2.1} which consists in multiplying $t$ by a positive constant can be
compensated for by dividing $a(r)$ by  the same constant. Thus modulo this coordinate
freedom one can say that the conditions $a \cdot b = 1$  and $\lambda_0=\lambda_1$ are 
equivalent.

If  $\lambda_0 = \lambda_1$, then   $\lambda_2$  simplifies from  eq. \eqref{2.5}  to
\begin{equation} 
\label{2.7}
\lambda_2 =  \frac{2aa'}{r} =  \frac{\left( a^2 \right)'}{r}.
\end{equation}
Assuming additionally $\lambda_2 = -2c$ to be constant, we get from eq. \eqref{2.7}
via $2rc =  -(a^2)'$  the solution for the metric components from  \eqref{2.1}
\begin{equation} 
\label{2.8}
a = \frac{1}{b} = \sqrt{ \tilde c - cr^2 }
\end{equation}
with a further constant $\tilde c$. Inserting eq. \eqref{2.8} into eqs. \eqref{2.3} and  \eqref{2.6} 
we get for this case $\lambda_0= - 2c$ and $R= - 6c$.  Therefore from only the two 
conditions  $\lambda_0 = \lambda_1$ and $\lambda_2 = {\rm  const.}$
we find that the space-time of  \eqref{2.1} is one of 
constant curvature i.e.  $\lambda_0 = \lambda_1 =\lambda_2 = {\rm const.}$
  
Eqs. \eqref{2.1} and \eqref{2.8} can be combined to give the metric 
\begin{equation} 
\label{2.9}
ds^2 = (\tilde c - cr^2) dt^2 - \frac{dr^2}{\tilde c - cr^2} - r^2 d \theta^2 .
\end{equation}
This space-time has a horizon at $r=r_h$ where $r_h$ is defined by 
$\tilde c = cr_h^2$. In contrast to the Schwarzschild metric in $3+1$ dimensions
the horizon for the space-time in \eqref{2.9} cannot be defined locally, since the character 
of the local isometry group of metric \eqref{2.9} does not change upon crossing the horizon. 
Furthermore the usual Schwarzschild solution in $3+1$ dimensions is only static outside the horizon, 
whereas the metric in \eqref{2.9} is static everywhere. The metric in \eqref{2.9} has more in common with de Sitter or
Anti de Sitter space-time. Another feature of the metric in  \eqref{2.9} is that the Ricci scalar is constant, 
$R= -6c$. Thus one is free to change the value of  $\tilde c$  by a suitable 
coordinate transformation. In order to satisfy eq. \eqref{2.2} one should set $\tilde c =1$. 

In case of $c=0$, the metric \eqref{2.9} is completely flat.  Nevertheless, a non-trivial 
feature can be constructed, if we change the character of the coordinate
$\theta$: if $\theta=0$ and $\theta = 2 \pi \mu$  are identified with
$0<\mu<1$, then we get a conical singularity at $r=0$ with a deficit angle
of  $2\pi(1-\mu)$. Although the space-time is locally flat there are still global 
effects such as the deflection of null geodesics. 

\section{Exact $2+1$ solution with real scalar field} 

In the preceding section we gave some general discussion of $2+1$ dimensional space-times. We now
use this discussion to write down a simple, exact solution with a self-interacting, real, scalar field coupled to 
the $2+1$ dimensional space-time. Since we are now considering sources we begin by writing Einstein's 
equations for $2+1$ dimensions
\begin{equation}
\label{3.1}
G_{\mu \nu} = \kappa T_{\mu \nu} ~,
\end{equation}
where $\kappa = 8 \pi G_3$ is the $2+1$ dimensional gravitational coupling
and $G_3$ is Newton's constant in $2+1$ dimensions; $\kappa$ has dimensions of 
inverse mass in units where $c=1$. \footnote{In \cite{giddings}, the following argument 
is given: In $n$-dimensional space-times and  in units where $c=1$, the dimension of  
$\kappa$ is $ {\rm length}^{n-3}/{\rm mass}$, this  explains the special role of $n=3$.}  
We use the convention of Landau and Lifshitz \cite{ll}. The metric ansatz is a specialization
of eq. \eqref{2.1} with $a ^2 (r) \rightarrow A(r)$ and $b(r) \rightarrow 1$
\begin{equation}
\label{3.2}
ds^2 = A(r) dt^2 -dr^2 - r^2 d \theta^2 ~. 
\end{equation}
The more general form of the metric given in eq. \eqref{2.1} did not yield any (obvious)
exact, closed form solutions.
We first calculate $G_{\mu \nu}$ for  eq. \eqref{3.2}  to get the left hand side of 
the $2+1$ Einstein's equation \eqref{3.1}
\begin{equation}
\label{3.3}
G_{rr} = \frac{A'(r)}{2 r A(r)}~~;~~ G_{\theta \theta} = \frac{r^2}{4 A^2(r)} 
\left[ -(A'(r))^2 +2 A(r) A''(r) \right] ~.
\end{equation}
All other components of $G_{\mu \nu}$ are zero. One oddity is that $G_{tt} =0$ 
which implies, by Einstein's equation \eqref{3.1}, that this metric has a vanishing 
energy density, $T_{tt}$. However this is not a vacuum solution (which in $2+1$ 
dimensions would necessarily be flat, Minkowski space-time) since the space-time 
has non-zero pressures, due to the non-vanishing of $G_{rr}$ and 
$G_{\theta \theta}$ for general $a(r)$. From a $3+1$ perspective 
a source with no energy-density but a non-vanishing pressure is at the very least 
odd. However, we note in this regard that in the BTZ black hole solution, if 
one views the negative cosmological constant as a constant energy-momentum tensor --
$T^{BTZ} _{\mu \nu} \propto \Lambda g_{\mu \nu}$  with $\Lambda <0$ -- then one 
has, not a zero energy density, but a negative energy density. As we will see, the 
reason for the vanishing  of the energy-density of this solution comes from a 
cancellation of the kinetic and potential energies of the scalar field. This oddity 
points out the unusual and anti-intuitive (from a $3+1$ dimensional perspective)
relationship between sources and the gravitational field in $2+1$ dimensions. 

To this $2+1$ space-time from eq. \eqref{3.2} we couple a real scalar field, $\phi$, with the
Lagrangian
\begin{equation}
\label{3.4}
{\cal L}_S = \frac{1}{2} \partial _\mu \phi \partial ^\mu \phi - V(\phi)
= \frac{1}{2} \partial _\mu \phi \partial ^\mu \phi + \frac{1}{2 \kappa} 
e^{-2 \sqrt{\kappa} \phi} ~,
\end{equation}
where the potential implied by  eq. \eqref{3.4} is a Liouville potential i.e 
$$
V(\phi) = - \frac{1}{2 \kappa} e^{-2\sqrt{\kappa} \phi}.
$$  
The presence of a potential which is a transcendental function can be compared to the 
use of  the sinusoidal potential, $V_{SG} (\phi) \propto \sin (k \phi )$, in the sine-Gordon 
soliton \cite{SG-soliton}. In the context of dilatonic gravity in $2+1$ dimensions 
such an exponential potential was used \cite{mann} to find a $2+1$ dimensional, charged 
black hole. In addition such exponential potentials are widely used in 
cosmology in $3+1$ in quintessence explanations of dark  energy \cite{ferreira}.
Such exponential potentials occur also  generically for the dilaton field in string
theory \cite{gasperini} and  brane models.  Also one might expect (and we will find)
that a field in $2+1$ dimensions  would have logarithmic behavior for large $r$.  
Taking the potential to be an exponential of 
the field will undo the logarithmic form of $\phi$ and allow one to combine the potential 
term with the kinetic energy term $\frac{1}{2} \partial _\mu \phi \partial ^\mu \phi$. 

From eq. \eqref{3.4} we need to calculate the energy-momentum tensor, 
$T_{\mu \nu}$, so that together with  the Einstein tensor, $G_{\mu \nu}$, from eq. \eqref{3.3}  
we can set up the Einstein equation \eqref{3.1}. Since we are assuming only $r$ dependence
of the metric we make the same assumption  for the field, $\phi$ i.e. 
$\phi (t, r, \theta) \rightarrow \phi (r)$. The general expression
for $T_{\mu \nu}$ associated with eq. \eqref{3.4} is
\begin{equation} 
\label{3.5}
T_{\mu \nu} = \partial _\mu \phi \partial _\nu \phi - g_{\mu \nu} {\cal L}_S.
\end{equation}
Taking into account $\phi (t, r, \theta) \rightarrow \phi (r)$ and the diagonal form 
of the metric \eqref{3.2} we get
\begin{equation}
\label{3.6}
T_{tt} = - A(r) {\cal L}_S ~~~~~;~~~~~ T_{rr} = (\phi ')^2 + {\cal L}_S ~~~~~;~~~~~
T_{\theta \theta} = r^2 {\cal L}_S ~.
\end{equation}
Since $G_{tt}=0$ we need $T_{tt} =0$. This implies that ${\cal L}_S=0$ for the 
condition that the solution satisfies $\phi (t, r, \theta) \rightarrow \phi (r)$ i.e. we want
\begin{equation}
\label{3.7}
{\cal L}_S=0 \rightarrow - \frac{1}{2}(\partial _r \phi)^2 + \frac{1}{2\kappa}
e^{-2 \sqrt{\kappa} \phi} =0 \rightarrow \phi ' (r) = 
\pm \frac{1}{\sqrt{\kappa}}e^{- \sqrt{\kappa} \phi} ~.
\end{equation}
At the end we will need to verify that our solution, $\phi (r)$, does in fact satisfy 
$\phi ' (r) =  \pm g e^{- \sqrt{\kappa}\phi} / \sqrt{\kappa}$. Note that 
this condition also gives $T_{rr} = (\phi ')^2$ and $T_{\theta \theta}=0$. Thus the only
non-zero source term is the radial pressure, $T_{rr}$.  The condition given in eq. \eqref{3.7} 
has the physical meaning that the kinetic energy of the scalar field
(the derivative term) is balanced and canceled by the potential energy of 
the scalar field (the exponential term). This balancing of kinetic and potential 
energies of $\phi$ makes 
it easy to understand the vanishing of the energy density, $T_{tt}$. 

Putting the $G_{\mu \nu}$'s from eq. \eqref{3.3} together with 
the $T_{\mu \nu}$'s from eq. \eqref{3.5} we arrive at the system of equation we need to solve
\begin{equation} 
\label{3.8}
\frac{A'(r)}{2 r A(r)} = \kappa (\phi ')^2 ~~~~~~;~~~~~ 
\frac{r^2}{4 A^2(r)} \left[ -(A'(r))^2 +2 A(r) A''(r) \right] =0 ~.
\end{equation}
In addition to finding an $A(r)$ and $\phi (r)$ which solve eq. \eqref{3.8} we also need 
$\phi (r)$ to solve the  Klein-Gordon equation coming from the Lagrange 
density  eq. \eqref{3.4} and the metric eq. \eqref{3.2}. In general,
the Klein-Gordon equation for $\phi$ in a curved background is
\begin{equation}
\label{3.9}
\frac{1}{\sqrt{g}} \partial _\mu \left( \sqrt{g} g^{\mu \nu} \partial_\nu \phi \right) 
= - \frac{\partial V}{\partial \phi} ~,
\end{equation}
where $\sqrt{g}=\sqrt{\det(g_{\mu \nu})}= r\sqrt{A(r)}$, and $g$ is the determinant of the
$2+1$ dimensional metric. Taking into account that $\phi (t, r, \theta) \rightarrow \phi (r)$
and $V(\phi) = -\frac{1}{2 \kappa} e^{-2 \sqrt{\kappa} \phi}$ 
we find that the Klein-Gordon equation for $\phi$ becomes
\begin{equation}
\label{3.10}
\phi '' (r) + \left( \frac{1}{r} + \frac{A'(r)}{2 A(r)} \right) \phi ' (r) = 
\frac{1}{\sqrt{\kappa}} e^{-2 \sqrt{\kappa} \phi} ~,
\end{equation}
where the primes are differentiation with respect to $r$. It is straightforward to show 
that the system of Einstein equation eq. \eqref{3.8} plus Klein-Gordon equation eq. \eqref{3.10}
are solved via
\begin{equation}
\label{3.11}
A(r) = Kr^2 ~~~~~; ~~~~~ \phi (r) = 
\frac{1}{\sqrt{\kappa}} \ln \left( r \right) ~.
\end{equation}
Finally, recall that in order for this solution in eq. \eqref{3.11} to work it must also 
satisfy the condition $\phi ' (r) = \pm  e^{- \sqrt{\kappa}\phi} / \sqrt{\kappa}$. It 
is straightforward to check that eq. \eqref{3.11} does satisfy this condition. Note that since
the only non-trivial metric component is, $g_{00} = K r^2$, one could absorb the 
constant $K$ into a redefinition of the time as $t \rightarrow t'/\sqrt{K}$. We keep
the constant $K$ for later comparison between the metric given by  eq. \eqref{3.2}  plus
$A(r) = K r^2$ and the asymptotic form of the AdS metric as well as the metrics of reference \cite{beato}. 

\section{Conclusions and Summary}

In this article we have found an $r$-symmetric, exact, analytical solution for a real 
scalar field coupled to gravity in $2+1$ dimensions. The solution is given by the metric 
of the form \eqref{3.2} with the ansatz function $A(r)$ and the scalar field, $\phi (r)$ 
given in eq. \eqref{3.11}. Such static solutions for gravitating real scalar fields in $2+1$ 
dimensions have received very little attention outside of a few works \cite{beato}. The 
solution presented here can be considered a $2+1$ dimensional version of the solutions
in $3+1$ dimensions of gravitating real scalar fields \cite{kodama}, \cite{baekler}, 
\cite{schmo}, and  \cite{jetzer} but having features that are unique to $2+1$ dimensions. 

The $\ln (r)$ behavior of the scalar field is what one expects for a massless field 
in $2+1$ dimensions in the absence of gravity e.g. in $2+1$ electrodynamics the scalar 
potential for a point charge goes as $\ln (r)$. Because of this one might think that the 
space-time given by  eq. \eqref{2.2} with $A(r) = K r^2$ is flat. Calculating scalar 
quantities such as the Ricci scalar $R^\alpha _\alpha =2/r^2$ or the square of the
Ricci tensor $R^{\mu \nu} R_{\mu \nu} = 2/r^4$ one finds that these scalars are not
zero for finite $r$  as would be the case with Minkowski space-time. (In the limit
$r \rightarrow \infty$ these scalars do vanish).  One can also see from these scalar quantities 
that the solution appears to possess a real singularity at $r=0$. This can be further confirmed
by calculating the trace of the energy-momentum tensor -- $T= T^\mu _\mu = T^r _r = -\frac{1}{\kappa r^2}$. 
Further more one can see that this is a naked
singularity since the solution given in eq. \eqref{3.11} does not have a horizon. Given the naked
singularity at $r=0$ and the $\ln (r)$ behavior of the scalar field one physical interpretation of 
this exact solution is that it represents the field of a self gravitating scalar field/particle
in $2+1$ dimensions. However, one should note that the existence of this solution depends 
also on the exponential self-interaction (i.e. Liouville potential).

In the spatial coordinates, $(r, \theta)$ the metric is asymptotically flat. In other 
words as $r \rightarrow \infty$ the  spatial part of the metric eq. \eqref{3.2} approaches
the 2D Euclidean line element $dr^2 +r^2 d \theta ^2$. The temporal part of the metric 
goes as $K r^2 dt^2$ which is equivalent to the asymptotic behavior of the temporal 
part of  Anti-de Sitter space-time  $\Lambda r^2 dt^2$ where $\Lambda$ is a positive
cosmological constant. This feature of the metric -- the spatial part of the metric being 
flat while the temporal part is equivalent to the asymptotic form of AdS space-time or 
BTZ space-time -- is connected with the unusual feature that the curvature of the 
space-time comes entirely from the radial pressure, $T_{rr}$. 
Additionally we can compare the metric of our solution 
$ds^2 = K r^2 dt^2 - dr^2 - r^2 d \theta ^2$ with some of
the static metric solutions discussed by E. Ay{\'o}n-Beato et al. \cite{beato} who 
found metric solutions of the form 
$$
ds^2 = (r^2/l^2) dt^2 - (l^2/r^2) dr^2 - r^2 d \theta ^2
$$
where $l$ is some length scale and $1/l^2$ can be taken as a cosmological constant.
In comparing our metric solution with this metric or with the asymptotic AdS metric, 
we see that the temporal parts have the same $r^2$ behavior but the $dr^2$ terms 
are different. Finally we note that our metric scales simply under the dilation  of the 
$r-$coordinate $r \rightarrow C r'$ where $C$ is some constant. Under this transformation
the metric simply scales like $C^2$ i.e. ${ds}^2 \rightarrow C^2 {ds'}^2$. 

As a closing note let us briefly mention that quite recently, several papers appeared 
which are related to different other aspects of the study of $(2+1)$-dimensional gravity:
In \cite{s25} and  \cite{s37}, massive gravity  and supergravity are discussed; 
in \cite{s29}, higher spin in topologically massive gravity is discussed;
in \cite{s26} and  \cite{s38} wormholes and star models are constructed 
in 2+1 dimensions;  in  \cite{s28},  Birkhoff's   theorem is generalized; 
in \cite{s31},  \cite{s32}, \cite{s33},  \cite{s34},  
\cite{s35}, \cite{s36},  \cite{s39}, and \cite{s27} various properties of 
the BTZ geometry are discussed; in \cite{s996}, massive particles with spin in $2+1$ dimension
are constructed; in \cite{s991}, \cite{s992}, \cite{s993}, \cite{s994}, and   
\cite{s995} further aspects of  $(2+1)$-dimensional gravity are deduced and discussed; and finally
in \cite{s998}, the observability of strong gravitational sources (black hole, naked singularities) 
via lensing is discussed.

\section*{Acknowledgments}
Useful comments by J. Barrow, S. Deser, M. Hortacsu, K.  Virbhadra, and
D.-H. Yeom  are gratefully acknowledged.  DS acknowledges the support of a
DAAD (Deutscher Akademischer Austauschdienst) grant to do research at
Universit{\"a}t Potsdam, and a 2012-2013 Fulbright Senior Scholar Grant to work at ITB.

\end{document}